\documentclass[letterpaper,english,letterpaper,twocolumn,aps,prb,superscriptaddress]{revtex4}

\usepackage{amsmath}
\usepackage{graphicx}

\makeatletter
\renewcommand\[{\begin{equation}} 

\renewcommand\]{\end{equation}} 

\usepackage{babel}
\makeatother

\begin{document}
\newcommand{\avg}[1]{\langle#1\rangle}

\newcommand{\p}{\prime}

\newcommand{\dg}{\dagger}

\newcommand{\ket}[1]{|#1\rangle}

\newcommand{\bra}[1]{\langle#1|}

\newcommand{\proj}[2]{|#1\rangle\langle#2|}

\newcommand{\inner}[2]{\langle#1|#2\rangle}

\newcommand{\tr}{\mathrm{tr}}

\newcommand{\pd}[2]{\frac{\partial#1}{\partial#2}}

\newcommand{\der}[2]{\frac{d #1}{d #2}}

\newcommand{\im}{\imath}

\renewcommand{\onlinecite}[1]{\cite{#1}}

\title{Finite representations of continuum environments}

\author{Michael Zwolak}

\email{mpz@lanl.gov}

\affiliation{Theoretical Division, MS-B213, Los Alamos National Laboratory, Los
Alamos, New Mexico 87545}

\affiliation{Institute for Quantum Information, California Institute of Technology,
Pasadena, California 91125}

\date{\today{}}

\begin{abstract}
Understanding dissipative and decohering processes is fundamental
to the study of quantum systems. An accurate and generic method for
investigating these processes is to simulate both the system \emph{and}
environment, which, however, is computationally very demanding. We
develop a novel approach to constructing finite representations of
the environment based on the influence of different frequency scales
on the system's dynamics. As an illustration, we analyze a solvable
model of an optical mode decaying into a reservoir. The influence
of the environment modes is constant for small frequencies, but drops
off rapidly for large frequencies, allowing for a very sparse representation
at high frequencies that gives a significant computational speedup
in simulating the environment. This approach provides a general framework
for simulating open quantum systems.
\end{abstract}


\maketitle
How quantum systems behave when in contact with an environment has
been the subject of a tremendous amount of research \citep{Weiss93-1,Leggett87-1,Zurek03-1}.
Many interesting physical scenarios can be cast in this form, such
as transport through nanoscale systems \citep{Chen05-1}, interaction
between local magnetic moments and conduction electrons in the Kondo
problem \citep{Kirchner08-1}, charge transfer in biomolecules \citep{Zwolak08-4},
decoherence of qubits \citep{Merkli07-1}, relaxation and decay \citep{Longhi06-1},
dissipative quantum phase transitions \citep{Meidan07-1}, and the
quantum-to-classical transition \citep{Zurek03-1}. Techniques to
study ``open'' systems thus have a wide range of applicability and
influence. 

Analytical techniques to study these systems are generally based on
integrating out the environment \citep{Keldysh65-1,Kadanoff62-1,Feynman63-1}.
However, to find solutions more often than not one has to resort to
a series of uncontrolled approximations. Numerical techniques, then,
are the key to accurate results for many physical systems. On the
one hand, Monte Carlo simulations can be used to calculate system
properties directly from the path-integral representation where the
environment has been integrated out \citep{Werner05-1}. On the other
hand, the numerical renormalization group (NRG) approach is based
on simulating the system and environment by choosing a finite representation
of the continuum environment \citep{Wilson75-1,Bulla07-1}. This technique
uses a logarithmic discretization of the environment's spectral density
\citep{Wilson75-1,Bulla07-1,Campo05-1}, which enforces a flow of
the low energy spectrum as one successively incorporates lower energy
degrees of freedom of the environment. However, since a variational
matrix-product state (MPS) approach \citep{Verstraete05-1} does not
require separating energy scales, one can ask a very fundamental question:
how do different fractions of the environment influence the system's
dynamics and how can this be used to construct efficient representations
of the environment?

\begin{figure}
\begin{centering}
\includegraphics[width=8cm]{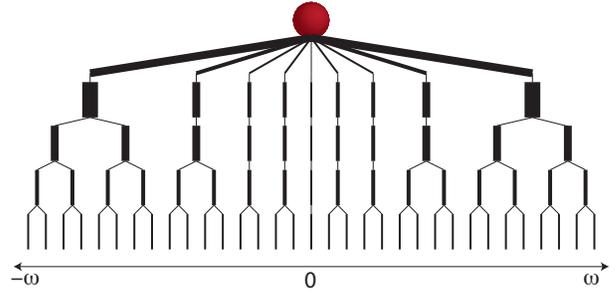}
\par\end{centering}

\caption{Schematic of a system connected to an environment. The system (top
sphere) is connected to a continuum, or very finely spaced, environment,
as shown on the lowest level. However, high frequency modes have less
influence over the system's dynamics, allowing for a very sparse representation
by grouping modes together. \label{fig:LevelGrouping}}

\end{figure}

We get insight into the answer by developing a novel approach for
constructing a finite representation of the environment based on the
influence of environment modes (hereon just referred to as \emph{modes})
at\emph{ }different frequency scales. We illustrate this strategy
by examining a solvable model of an optical cavity mode decaying into
a reservoir, where we can obtain exactly the dynamics of the system
connected to a discrete environment composed of evenly spaced modes,
see Fig. \eqref{fig:LevelGrouping}. We show that there is a frequency
window centered around the system's frequency where the mode influence
is constant. However, outside this window, the mode influence drops
off as $\left(\Delta/\omega^{2}\right)^{2}$, where $\Delta$ is the
mode spacing and $\omega$ is the mode frequency. Thus, instead of
using evenly spaced modes, we use an alternative, frequency dependent
discretization $\Delta\left(\omega\right)=\Delta_{o}+d\omega^{2}$
that enforces the influence of modes at large frequencies to be constant.
With this discretization, the computational cost of the simulation
is significantly reduced.

We begin with a general description of the problem. A generic Hamiltonian
of a system connected to an environment is \begin{equation}
H=H_{S}+\sum_{k}\left(g_{k}L_{S}^{\dg}b_{k}+g_{k}^{\star}b_{k}^{\dagger}L_{S}\right)+\sum_{k}\omega_{k}b_{k}^{\dagger}b_{k}\label{eq:genericH}\end{equation}
where $H_{S}$ is the system Hamiltonian and $L_{S}$ is some system
operator. This represents a system connected to a non-interacting
set of environment modes linearly in the environment operators. For
this type of connection, the spectral function \begin{equation}
J\left(\omega\right)=\sum_{k}\left|g_{k}\right|^{2}\delta\left(\omega-\omega_{k}\right)\label{eq:spectral}\end{equation}
completely defines the couplings and modes of the environment. Typically
the spectral function is taken as some continuous function of frequency
to indicate that for all practical purposes the environment is infinite
compared to the system. To make simulating the system and environment
a viable technique, one needs a controlled procedure for performing
the mapping $J\left(\omega\right)\to\left\{ \left(\omega_{i},G_{i}\right)\right\} $,
where the set of environment modes $\left\{ \left(\omega_{i},G_{i}\right)\right\} $
$ $ is finite and each mode $\omega_{i}$ has the associated coupling
constant $G_{i}=\sqrt{J\left(\omega_{i}\right)\Delta\left(\omega_{i}\right)}$.
Our starting point to find an efficient mapping is to define a measure
of error for an approximation of the environment. Quite generally
one is interested in system properties only, we thus use the error
measure \begin{equation}
\epsilon_{N}=\int_{0}^{T}dt\:\tr\left[\rho_{N}\left(t\right)-\rho_{ex}\left(t\right)\right]^{2}\label{eq:errormeasure}\end{equation}
where $T$ is the simulation time, $\rho_{N}$ is the reduced density
matrix of the system in the presence of the finite representation
(of size $N$) of the environment, and $\rho_{ex}$ is the exact reduced
density matrix of the system in the presence of the continuum environment
\footnote{Alternative measures are appropriate if one is interested in a particular
observable or the relative error.%
}. We also define a measure of mode influence as\begin{equation}
I_{N/\omega}=\int_{0}^{T}dt\:\tr\left[\rho_{N/\omega}\left(t\right)-\rho_{N}\left(t\right)\right]^{2}\label{eq:influence}\end{equation}
where $\rho_{N/\omega}$ is the reduced density matrix in the presence
of all but one, at frequency $\omega$, of the $N$ modes. The intuition
behind using Eq. \eqref{eq:influence} is that modes with a small
influence should be removable in a controllable manner %
\footnote{Since we plan to remove many modes, the spectral density of these
modes must be included with the remaining modes. Another question
one can ask is if we remove two modes and replace them with one, what
error is incurred? This error has similar behavior to the influence
in Eq. \eqref{eq:influence}.%
}. Having a guide such as Eq. \eqref{eq:influence} is crucial because
an efficient choice of modes is going to be dependent on many factors
- whether one wants equilibrium or real-time properties, the time/temperature
of the simulation, the initial conditions (such as the initial excitation
of the system and temperature of the environment), the Hamiltonian
under consideration, etc. As a case in point, recently we showed that
environments which give polynomially decaying memory kernels naturally
motivate a logarithmic discretization of the kernel in time, and a
property called \emph{increasing-smoothness} provides a guide for
more generally determining an efficient discretization \citep{Zwolak08-1}.

To further describe and illustrate the approach, let us analyze an
example of a single optical cavity mode \citep{Carmichael93-1} decaying
into a reservoir with the Hamiltonian %
\footnote{We remove the conserved term $\omega_{o}\left(a^{\dg}a+\sum_{k}b_{k}^{\dg}b_{k}\right)$
and take the system frequency $\omega_{o}$ to be large.%
} \begin{equation}
H=a^{\dagger}B+B^{\dg}a+\sum_{k}\omega_{k}b_{k}^{\dg}b_{k}\label{eq:decayH}\end{equation}
where $B=\sum_{k}g_{k}b_{k}$. We consider the initial state defined
by the correlation functions $\left\langle a^{\dg}a\right\rangle =1$
and $\left\langle b_{k}^{\dg}b_{k}\right\rangle =0$, and follow the
evolution of the number of particles in the system, $n\left(t\right)=\left\langle a^{\dg}a\right\rangle _{t}$,
which determines the reduced density matrix of the system by $\rho\left(t\right)=\mathrm{diag}\left[n\left(t\right),1-n\left(t\right)\right]$.
We also take a constant spectral function $J\left(\omega\right)=\frac{\gamma}{2\pi}$
. The exact solution in the continuum limit is $n\left(t\right)=e^{-\gamma t}$.
The nice property of this model, however, is that we can solve exactly
the system dynamics in the presence of discrete, evenly spaced modes
\citep{Milonni83-1}. The solution for the system connected to an
infinite set of modes with spacing $\Delta$ is \[
n\left(t\right)=e^{-\gamma t}+\Theta\left(t-\frac{2\pi}{\Delta}\right)F\left(t,\Delta\right)\]
where $\Theta\left(t-\frac{2\pi}{\Delta}\right)$ is the step-function
and the function $F\left(t,\Delta\right)$ is a sum of correction
terms of order one. That is, for an infinite set of evenly spaced
modes, the dynamics of the system is exact up to time $T=2\pi/\Delta$.
Based on this nice property we can now examine two approaches to constructing
a finite representation of the environment.

\emph{Linear Discretization }- Since the simulation is exact for evenly
spaced modes with $\Delta\le2\pi/T$, the error incurred in constructing
a finite representation is going to be due to the choice of cutoff
$\omega_{c}$. In the continuum limit, imposing a cutoff gives an
error, from Eq. \eqref{eq:errormeasure}, \begin{equation}
\epsilon_{c}=\frac{5}{2\gamma}\left(\frac{\gamma}{\pi\omega_{c}}\right)^{2}\approx\frac{5}{2\gamma}\left(\frac{\gamma T}{\pi^{2}N}\right)^{2}\label{eq:cutofferror}\end{equation}
to highest order in $1/\omega_{c}$ and the second expression is with
the largest spacing possible to get a controllable error, $\Delta=2\pi/T$.
This gives a direct approach to constructing a finite representation
of the environment. Given the simulation time $T$, one simply uses
a mode spacing of $2\pi/T$ and uses the bandwidth $\left[-\omega_{c},\omega_{c}\right]$
as a control parameter to approach the exact dynamics within time
$T$. We will see below this error behavior in comparison to an alternative,
influence discretization (ID).

\begin{figure}
\begin{centering}
\includegraphics[width=8cm]{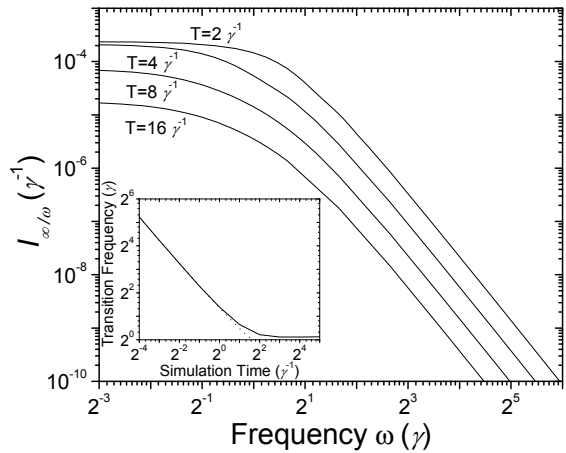}
\par\end{centering}

\caption{Single mode influence versus frequency for several simulation times.
For large frequencies the influence drops off as $\left(\Delta/\omega^{2}\right)^{2}$
and for small frequencies it is constant within some frequency window.
The inset shows the size of this frequency window, which drops off
as $1/T$ for short times and approaches a constant for long times.\label{fig:influenceplot}}

\end{figure}

\emph{Influence Discretization} - The previous approach treats all
of the modes within the bandwidth on equal footing, i.e., they all
are coupled with the same strength to the system. However, it does
implicitly recognize that high-frequency modes matter less, and can
be truncated with a controllable error. Thus, we want to develop a
way that explicitly recognizes that some frequency scales matter less,
i.e., have a smaller influence on the system dynamics. To do this,
let us solve for the influence, $I_{\infty/\omega}$. In the presence
of an infinite bandwidth of evenly spaced modes, the equation of motion
for $a\left(t\right)$, valid up to time $T=2\pi/\Delta$, is \[
\dot{a}\left(t\right)=-\gamma a\left(t\right)/2\,.\]
If we remove a single mode at frequency $\omega$, the equation of
motion becomes \[
\dot{a}\left(t\right)=-\gamma a\left(t\right)/2+\frac{\gamma\Delta}{2\pi}\int_{0}^{t}dt^{\p}\, e^{-i\omega\left(t-t^{\p}\right)}a\left(t^{\p}\right)\,,\]
which is again valid up to time $T=2\pi/\Delta$ and the constant
$\gamma\Delta/2\pi\equiv G_{\omega}^{2}$ is the square of the coupling
constant to mode $\omega$. The solution is \[
a\left(t\right)=\frac{s_{+}+i\omega}{s_{+}-s_{-}}e^{s_{+}t}-\frac{s_{-}+i\omega}{s_{+}-s_{-}}e^{s_{-}t}\,,\]
where $s_{\pm}=-\frac{1}{2}\left(\frac{\gamma}{2}+i\omega\right)\pm\frac{1}{2}\sqrt{\sigma}$
and $\sigma=\left(\frac{\gamma}{2}\right)^{2}+4\frac{\gamma\Delta}{2\pi}-\omega^{2}-i\omega\gamma$.
Since we have the time evolution of the operator $a\left(t\right)$,
we can now directly calculate the influence in Eq. \eqref{eq:influence}.
We plot this influence as a function of frequency for several times
in Figure \eqref{fig:influenceplot}. 

There are two crucial observations: (\emph{i}) the influence is constant
for small frequencies within a frequency window of width proportional
to $1/T$, and (\emph{ii}) for large frequencies the influence drops
off as $\left(\Delta/\omega^{2}\right)^{2}$. This suggests an uneven
spacing of modes that is approximately constant at small frequencies
and switches over to $ $a spacing proportional to $\omega^{2}$ for
large frequencies. The simplest mode spacing %
\footnote{The spectral function appears with $\Delta$ as the product $J\left(\omega\right)\Delta$,
suggesting a more general spacing $\Delta_{o}+d\omega^{2}/J\left(\omega\right)$.%
} that has this behavior is \begin{equation}
\Delta\left(\omega\right)=\Delta_{o}+d\omega^{2}\,.\label{eq:spacing}\end{equation}
This spacing results in a constant influence proportional to $d^{2}$
at large frequencies %
\footnote{We set $\Delta_{o}=2\pi/2T$ since low frequency modes will give large
recurrence errors if their spacing is larger than $2\pi/T$$ $.%
}. It also enables the treatment of both the truncated modes and other
high frequency modes on a similar footing. That is, beyond a frequency
$\omega_{c}\propto1/d$ one can not choose any more modes and thus
the parameter $d$ defines a natural cutoff, which, as we decrease
$d$, we both increase the number of modes that are approximately
evenly spaced at low frequency and increase the cutoff. From Eq. \ref{eq:cutofferror},
the error due to the cutoff, which corresponds to its influence, is
also proportional to $d^{2}$. Thus, the choice of mode spacing \eqref{eq:spacing}
enforces the high-frequency modes to be treated equally based on influence
rather than based on coupling.

\begin{figure}
\begin{centering}
\includegraphics[width=8cm]{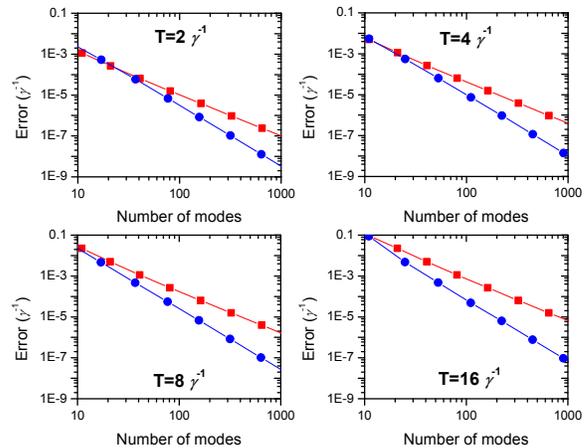}
\par\end{centering}

\caption{Simulation results for the error (Eq. \eqref{eq:errormeasure}) versus
the number of modes for several simulation times. For all the times,
the evenly spaced approach (squares) gives an error that behaves as
$1/N^{2}$, whereas using the ID (circles) gives $1/N_{ID}^{3}$.
\label{fig:ErrorVsTime}}

\end{figure}

We can also obtain an expression for the error behavior of choosing
a number of modes $N_{ID}$ based on Eq. \eqref{eq:spacing}. A spacing
given by Eq. \eqref{eq:spacing} results in $N_{ID}\approx\int_{-\omega_{c}}^{\omega_{c}}d\omega/\Delta\left(\omega\right)\propto\sqrt{N}$
modes between $-\omega_{c}$ and $\omega_{c}$ ($\propto1/d$), compared
with $N$ evenly spaced modes for the same frequency cutoff. In the
process of sparsifying the high frequency modes (see Fig. \eqref{fig:LevelGrouping}),
$\mathcal{O}\left(\sqrt{N}\right)$ groupings of modes have to be
performed, each with an error $\mathcal{O}\left(d^{2}\right)$, which
are of the same order as the truncation error. Thus, the error of
the ID is $\epsilon_{ID}\propto\sqrt{N}\epsilon_{c}$ . Starting from
the error of the evenly spaced discretization, $\epsilon_{c}\propto1/N^{2}$,
one obtains \begin{equation}
\epsilon_{ID}\propto1/N_{ID}^{3}\,.\label{eq:ADerror}\end{equation}
We show the results of simulations in Fig. \eqref{fig:ErrorVsTime}
\footnote{An additional, numerical comparison with $\Delta\left(\omega\right)=\Delta_{o}+d\omega$
shows that the spacing \eqref{eq:spacing} is more efficient.%
}. We obtain the error scalings from Eqs. \eqref{eq:cutofferror} and
\eqref{eq:ADerror}, and we further see that except for very short
simulation times and very small number of modes, the ID needs significantly
fewer modes to achieve the same accuracy. In these simulations, the
computational cost to simulate an environment of $N$ modes is $N^{3}$.
Thus, given a desired error, the ID reduces the computational cost
from $N^{3}$ to $N_{ID}^{3}=N^{2}$.

\emph{Beyond solvable models} - The above approach provides a general
framework for simulating {}``open,'' many-body Hamiltonians. Ultimately
one wants to have an efficient, finite representation of the environment
and a controllable procedure for taking the continuum limit. Therefore,
we suggest taking the general Hamiltonian \eqref{eq:genericH} and
first examining a solvable version of it, e.g., either take the quadratic
part or the part of $H_{S}$ that commutes with the system-environment
interaction. With this solvable version, one can compute exactly the
mode influence under similar conditions (dynamics/temperatures) to
be studied with the full $H_{S}$. Then use this to construct a discretization
and continuum limit procedure that will be used in the many-body case
\footnote{We conjecture that including a time-dependent field within the solvable
model will mimic the many-body interaction by creating a range of
accessible states. Likewise, we stress that the spacing is going to
be dependent on the particulars of the system, including how the environment
modifies the system, e.g., whether it shifts energies, mixes states,
etc.%
}. The latter ensures that even if the discretization is not as efficient,
there will still be control over the errors. Then, as with NRG, one
can use the Wilson-chain construction \citep{Wilson75-1} directly
with MPS simulations \citep{Vidal03-2,Vidal04-1,Zwolak04-1,Verstraete04-1},
where one can deal with both evenly or unevenly spaced modes due to
the variational nature of the MPS algorithms \citep{Verstraete05-1}.
This will be discussed in more detail in a later publication.

\emph{Conclusions} - We have developed a novel approach to constructing
finite representations of continuum environments by using the single-mode
influence \eqref{eq:influence} as a guide to choosing a distribution
of modes. In an illustrative case, we found that the influence drops
off as $\left(\Delta/\omega^{2}\right)^{2}$, which suggests a mode
spacing $\Delta\left(\omega\right)=\Delta_{o}+d\omega^{2}$ . This
treats the high-frequency modes, including the truncation of modes
beyond the artificially imposed cutoff, on a similar footing. For
this case %
\footnote{There will be a corresponding reduction in computational cost for
MPS simulations that is harder to assess since the effect on the MPS
dimension is unclear.%
}, we showed that the computational cost to achieve a given error is
reduced from $N^{3}$ to $N_{ID}^{3}=N^{2}$. It may be possible to
further increase the computational efficiency by adding Markovian
reservoirs to the environmental modes, where in some cases this exactly
replicates the continuous environment \citep{Garraway97-2},
or by including classical degrees of freedom in the dynamics \citep{Sergi05-1,Sergi07-1}.

\begin{acknowledgments}
We thank W. Zurek, G. Refael, G. Smith, F. M. Cucchietti, and P. Milonni
for helpful comments. This research was supported in part by a Gordon
and Betty Moore Fellowship at Caltech and by the U.S. Department of
Energy through the LANL/LDRD Program.
\end{acknowledgments}
\bibliographystyle{apsrev}
\bibliography{open}

\begin{thebibliography}{27}
\expandafter\ifx\csname natexlab\endcsname\relax\def\natexlab#1{#1}\fi
\expandafter\ifx\csname bibnamefont\endcsname\relax
  \def\bibnamefont#1{#1}\fi
\expandafter\ifx\csname bibfnamefont\endcsname\relax
  \def\bibfnamefont#1{#1}\fi
\expandafter\ifx\csname citenamefont\endcsname\relax
  \def\citenamefont#1{#1}\fi
\expandafter\ifx\csname url\endcsname\relax
  \def\url#1{\texttt{#1}}\fi
\expandafter\ifx\csname urlprefix\endcsname\relax\def\urlprefix{URL }\fi
\providecommand{\bibinfo}[2]{#2}
\providecommand{\eprint}[2][]{\url{#2}}

\bibitem[{\citenamefont{Weiss}(1993)}]{Weiss93-1}
\bibinfo{author}{\bibfnamefont{U.}~\bibnamefont{Weiss}},
  \emph{\bibinfo{title}{Quantum Dissipative Systems}}
  (\bibinfo{publisher}{World Scientific Publishing}, \bibinfo{year}{1993}).

\bibitem[{\citenamefont{Leggett et~al.}(1987)\citenamefont{Leggett,
  Chakravarty, Dorsey, Fisher, Garg, and Zwerger}}]{Leggett87-1}
\bibinfo{author}{\bibfnamefont{A.~J.} \bibnamefont{Leggett}},
  \bibinfo{author}{\bibfnamefont{S.}~\bibnamefont{Chakravarty}},
  \bibinfo{author}{\bibfnamefont{A.~T.} \bibnamefont{Dorsey}},
  \bibinfo{author}{\bibfnamefont{M.~P.~A.} \bibnamefont{Fisher}},
  \bibinfo{author}{\bibfnamefont{A.}~\bibnamefont{Garg}}, \bibnamefont{and}
  \bibinfo{author}{\bibfnamefont{W.}~\bibnamefont{Zwerger}},
  \bibinfo{journal}{Rev. Mod. Phys.} \textbf{\bibinfo{volume}{59}},
  \bibinfo{pages}{1} (\bibinfo{year}{1987}).

\bibitem[{\citenamefont{Zurek}(2003)}]{Zurek03-1}
\bibinfo{author}{\bibfnamefont{W.~H.} \bibnamefont{Zurek}},
  \bibinfo{journal}{Rev. Mod. Phys.} \textbf{\bibinfo{volume}{75}},
  \bibinfo{pages}{715} (\bibinfo{year}{2003}).

\bibitem[{\citenamefont{Chen and Di~Ventra}(2005)}]{Chen05-1}
\bibinfo{author}{\bibfnamefont{Y.-C.} \bibnamefont{Chen}} \bibnamefont{and}
  \bibinfo{author}{\bibfnamefont{M.}~\bibnamefont{Di~Ventra}},
  \bibinfo{journal}{Phys. Rev. Lett.} \textbf{\bibinfo{volume}{95}},
  \bibinfo{pages}{166802} (\bibinfo{year}{2005}).

\bibitem[{\citenamefont{Kirchner and Si}(2008)}]{Kirchner08-1}
\bibinfo{author}{\bibfnamefont{S.}~\bibnamefont{Kirchner}} \bibnamefont{and}
  \bibinfo{author}{\bibfnamefont{Q.}~\bibnamefont{Si}}, \bibinfo{journal}{Phys.
  Rev. Lett.} \textbf{\bibinfo{volume}{100}}, \bibinfo{pages}{026403}
  (\bibinfo{year}{2008}).

\bibitem[{\citenamefont{Zwolak and Di~Ventra}(2008)}]{Zwolak08-4}
\bibinfo{author}{\bibfnamefont{M.}~\bibnamefont{Zwolak}} \bibnamefont{and}
  \bibinfo{author}{\bibfnamefont{M.}~\bibnamefont{Di~Ventra}},
  \bibinfo{journal}{Rev. Mod. Phys.} \textbf{\bibinfo{volume}{80}},
  \bibinfo{pages}{141} (\bibinfo{year}{2008}).

\bibitem[{\citenamefont{Merkli et~al.}(2007)\citenamefont{Merkli, Sigal, and
  Berman}}]{Merkli07-1}
\bibinfo{author}{\bibfnamefont{M.}~\bibnamefont{Merkli}},
  \bibinfo{author}{\bibfnamefont{I.~M.} \bibnamefont{Sigal}}, \bibnamefont{and}
  \bibinfo{author}{\bibfnamefont{G.~P.} \bibnamefont{Berman}},
  \bibinfo{journal}{Phys. Rev. Lett.} \textbf{\bibinfo{volume}{98}},
  \bibinfo{pages}{130401} (\bibinfo{year}{2007}).

\bibitem[{\citenamefont{Longhi}(2006)}]{Longhi06-1}
\bibinfo{author}{\bibfnamefont{S.}~\bibnamefont{Longhi}},
  \bibinfo{journal}{Phys. Rev. Lett.} \textbf{\bibinfo{volume}{97}},
  \bibinfo{pages}{110402} (\bibinfo{year}{2006}).

\bibitem[{\citenamefont{Meidan et~al.}(2007)\citenamefont{Meidan, Oreg, and
  Refael}}]{Meidan07-1}
\bibinfo{author}{\bibfnamefont{D.}~\bibnamefont{Meidan}},
  \bibinfo{author}{\bibfnamefont{Y.}~\bibnamefont{Oreg}}, \bibnamefont{and}
  \bibinfo{author}{\bibfnamefont{G.}~\bibnamefont{Refael}},
  \bibinfo{journal}{Phys. Rev. Lett.} \textbf{\bibinfo{volume}{98}},
  \bibinfo{pages}{187001} (\bibinfo{year}{2007}).

\bibitem[{\citenamefont{Keldysh}(1965)}]{Keldysh65-1}
\bibinfo{author}{\bibfnamefont{L.~V.} \bibnamefont{Keldysh}},
  \bibinfo{journal}{Sov. Phys. JETP} \textbf{\bibinfo{volume}{20}},
  \bibinfo{pages}{1018} (\bibinfo{year}{1965}).

\bibitem[{\citenamefont{Kadanoff and Baym}(1962)}]{Kadanoff62-1}
\bibinfo{author}{\bibfnamefont{L.~P.} \bibnamefont{Kadanoff}} \bibnamefont{and}
  \bibinfo{author}{\bibfnamefont{G.}~\bibnamefont{Baym}},
  \emph{\bibinfo{title}{Quantum Statistical Mechanics}} (\bibinfo{publisher}{W.
  A. Benjamin, Inc.}, \bibinfo{address}{New York}, \bibinfo{year}{1962}).

\bibitem[{\citenamefont{Feynman and Vernon}(1963)}]{Feynman63-1}
\bibinfo{author}{\bibfnamefont{R.~P.} \bibnamefont{Feynman}} \bibnamefont{and}
  \bibinfo{author}{\bibfnamefont{F.~L.} \bibnamefont{Vernon}},
  \bibinfo{journal}{Ann. Phys. - New York} \textbf{\bibinfo{volume}{24}},
  \bibinfo{pages}{118} (\bibinfo{year}{1963}).

\bibitem[{\citenamefont{Werner and Troyer}(2005)}]{Werner05-1}
\bibinfo{author}{\bibfnamefont{P.}~\bibnamefont{Werner}} \bibnamefont{and}
  \bibinfo{author}{\bibfnamefont{M.}~\bibnamefont{Troyer}},
  \bibinfo{journal}{Phys. Rev. Lett.} \textbf{\bibinfo{volume}{95}},
  \bibinfo{pages}{060201} (\bibinfo{year}{2005}).

\bibitem[{\citenamefont{Wilson}(1975)}]{Wilson75-1}
\bibinfo{author}{\bibfnamefont{K.~G.} \bibnamefont{Wilson}},
  \bibinfo{journal}{Rev. Mod. Phys.} \textbf{\bibinfo{volume}{47}},
  \bibinfo{pages}{773} (\bibinfo{year}{1975}).

\bibitem[{\citenamefont{Bulla et~al.}(2008)\citenamefont{Bulla, Costi, and
  Pruschke}}]{Bulla07-1}
\bibinfo{author}{\bibfnamefont{R.}~\bibnamefont{Bulla}},
  \bibinfo{author}{\bibfnamefont{T.~A.} \bibnamefont{Costi}}, \bibnamefont{and}
  \bibinfo{author}{\bibfnamefont{T.}~\bibnamefont{Pruschke}},
  \bibinfo{journal}{Rev. Mod. Phys.} \textbf{\bibinfo{volume}{80}},
  \bibinfo{pages}{395} (\bibinfo{year}{2008}).

\bibitem[{\citenamefont{Campo and Oliveira}(2005)}]{Campo05-1}
\bibinfo{author}{\bibfnamefont{V.~L.} \bibnamefont{Campo}} \bibnamefont{and}
  \bibinfo{author}{\bibfnamefont{L.~N.} \bibnamefont{Oliveira}},
  \bibinfo{journal}{Phys. Rev. B} \textbf{\bibinfo{volume}{72}},
  \bibinfo{pages}{104432} (\bibinfo{year}{2005}).

\bibitem[{\citenamefont{Verstraete et~al.}(2005)\citenamefont{Verstraete,
  Weichselbaum, Schollw\"ock, Cirac, and von Delft}}]{Verstraete05-1}
\bibinfo{author}{\bibfnamefont{F.}~\bibnamefont{Verstraete}},
  \bibinfo{author}{\bibfnamefont{A.}~\bibnamefont{Weichselbaum}},
  \bibinfo{author}{\bibfnamefont{U.}~\bibnamefont{Schollw\"ock}},
  \bibinfo{author}{\bibfnamefont{J.~I.} \bibnamefont{Cirac}}, \bibnamefont{and}
  \bibinfo{author}{\bibfnamefont{J.}~\bibnamefont{von Delft}},
  \bibinfo{journal}{cond-mat/0504305}  (\bibinfo{year}{2005}).

\bibitem[{\citenamefont{Zwolak}(2008)}]{Zwolak08-1}
\bibinfo{author}{\bibfnamefont{M.}~\bibnamefont{Zwolak}},
  \bibinfo{journal}{cond-mat/0611412, to appear in Computational Science and
  Discovery}  (\bibinfo{year}{2008}).

\bibitem[{\citenamefont{Carmichael}(1993)}]{Carmichael93-1}
\bibinfo{author}{\bibfnamefont{H.~J.} \bibnamefont{Carmichael}},
  \emph{\bibinfo{title}{An Open Systems Approach to Quantum Optics}}
  (\bibinfo{publisher}{Springer-Verlag}, \bibinfo{address}{Berlin},
  \bibinfo{year}{1993}).

\bibitem[{\citenamefont{Milonni et~al.}(1983)\citenamefont{Milonni, Ackerhalt,
  Galbraith, and Shih}}]{Milonni83-1}
\bibinfo{author}{\bibfnamefont{P.~W.} \bibnamefont{Milonni}},
  \bibinfo{author}{\bibfnamefont{J.~R.} \bibnamefont{Ackerhalt}},
  \bibinfo{author}{\bibfnamefont{H.~W.} \bibnamefont{Galbraith}},
  \bibnamefont{and} \bibinfo{author}{\bibfnamefont{M.-L.} \bibnamefont{Shih}},
  \bibinfo{journal}{Phys. Rev. A} \textbf{\bibinfo{volume}{28}},
  \bibinfo{pages}{32} (\bibinfo{year}{1983}).

\bibitem[{\citenamefont{Vidal}(2003)}]{Vidal03-2}
\bibinfo{author}{\bibfnamefont{G.}~\bibnamefont{Vidal}},
  \bibinfo{journal}{Phys. Rev. Lett.} \textbf{\bibinfo{volume}{91}},
  \bibinfo{pages}{147902} (\bibinfo{year}{2003}).

\bibitem[{\citenamefont{Vidal}(2004)}]{Vidal04-1}
\bibinfo{author}{\bibfnamefont{G.}~\bibnamefont{Vidal}},
  \bibinfo{journal}{Phys. Rev. Lett.} \textbf{\bibinfo{volume}{93}},
  \bibinfo{pages}{040502} (\bibinfo{year}{2004}).

\bibitem[{\citenamefont{Zwolak and Vidal}(2004)}]{Zwolak04-1}
\bibinfo{author}{\bibfnamefont{M.}~\bibnamefont{Zwolak}} \bibnamefont{and}
  \bibinfo{author}{\bibfnamefont{G.}~\bibnamefont{Vidal}},
  \bibinfo{journal}{Phys. Rev. Lett.} \textbf{\bibinfo{volume}{93}},
  \bibinfo{pages}{207205} (\bibinfo{year}{2004}).

\bibitem[{\citenamefont{Verstraete et~al.}(2004)\citenamefont{Verstraete,
  Garcia-Ripoll, and Cirac}}]{Verstraete04-1}
\bibinfo{author}{\bibfnamefont{F.}~\bibnamefont{Verstraete}},
  \bibinfo{author}{\bibfnamefont{J.~J.} \bibnamefont{Garcia-Ripoll}},
  \bibnamefont{and} \bibinfo{author}{\bibfnamefont{J.~I.} \bibnamefont{Cirac}},
  \bibinfo{journal}{Phys. Rev. Lett.} \textbf{\bibinfo{volume}{93}},
  \bibinfo{pages}{207204} (\bibinfo{year}{2004}).

\bibitem[{\citenamefont{Garraway}(1997)}]{Garraway97-2}
\bibinfo{author}{\bibfnamefont{B.~M.} \bibnamefont{Garraway}},
  \bibinfo{journal}{Phys. Rev. A} \textbf{\bibinfo{volume}{55}},
  \bibinfo{pages}{4636} (\bibinfo{year}{1997}).

\bibitem[{\citenamefont{Sergi}(2005)}]{Sergi05-1}
\bibinfo{author}{\bibfnamefont{A.}~\bibnamefont{Sergi}},
  \bibinfo{journal}{Phys. Rev. E} \textbf{\bibinfo{volume}{72}},
  \bibinfo{pages}{066125} (\bibinfo{year}{2005}).

\bibitem[{\citenamefont{Sergi}(2007)}]{Sergi07-1}
\bibinfo{author}{\bibfnamefont{A.}~\bibnamefont{Sergi}}, \bibinfo{journal}{J.
  Phys. A: Math. Theor.} \textbf{\bibinfo{volume}{40}}, \bibinfo{pages}{F347}
  (\bibinfo{year}{2007}).

\end{thebibliography}

\end{document}